\documentclass[twocolumn,preprintnumbers,amsmath,amssymb]{revtex4}

\usepackage{epsfig}
\usepackage{dcolumn}
\usepackage{bm}
\usepackage{psfrag}

\newlength{\mywa}
\newlength{\myw}
\newcommand{\img}{\mathbf{i}}

\begin{document}
\preprint{CNL-2003}

%-----------------------------------------------------------------------------
\title{On the Propagation of Congestion Waves in the Internet}
\author{J\'ozsef St\'eger}%
\email{steger@complex.elte.hu}
\author{P\'eter Vaderna}
\altaffiliation[Also at ]{Traffic Analysis and Network Performance Laboratory\\
  Ericsson Hungary Ltd.
}
\email{Peter.Vaderna@eth.ericsson.se}
\author{G\'abor Vattay}\altaffiliation[Also at ]{Collegium Budapest, Institute for Advanced Study}
\email{vattay@complex.elte.hu}
\affiliation{
   Department of Physics of Complex Systems\\
   Eotvos University, Budapest \\
}%
\date{\today}
%-----------------------------------------------------------------------------

\begin{abstract}
Traffic modeling of communication networks such as Internet has become a very important field of research. A number of interesting phenomena are found 
in measurements and traffic simulations. One of them is the propagation
of congestion waves opposite to the main packet flow direction.
The purpose of this paper is to model and analyze packet congestion on
a given route and to give a possible explanation to this phenomenon.
\end{abstract}
\maketitle

%-----------------------------------------------------------------------------
% Introduction
%-----------------------------------------------------------------------------

In the recent past many aspects of computer data traffic on the Internet have been investigated. Signatures of long-range correlations\cite{4}, scaling\cite{1}, chaos\cite{2} and phase transitions\cite{3} have been found. In Ref.\cite{csab} $1/f^\alpha$ noise has been observed in the time series of round trip times similar to those observed in highway traffic measurements\cite{higuchi}. Since the slowing down and acceleration of packet flow in computer networks are very similar to those of cellular automaton models of cars in highway traffic\cite{5}, it has been argued that a valid analogy exists between these subjects which later has been demonstrated quantitatively\cite{6,Huisinga}. One of the most peculiar features of car traffic and flow of granular media is the propagation of density waves\cite{congestion}. In car traffic stop-and-go type congestion waves propagate against the direction of the flow. In Ref.\cite{taka} the authors made an attempt to observe the propagation of congestion in the routers of a real computer network. They studied the spatio-temporal correlations of the level of congestion in routers and showed that congestion can propagate from a heavily loaded router to one of its empty neighboring routers. Today computer networks can be studied with the help of network simulators, which has been developed by engineers.  These tools enable us to assemble any computer network configuration, to use the most commonly used packet sending mechanism TCP/IP protocol and to emulate its real behavior without building the system from hardware components. These simulators can imitate the behavior of hardware elements (computers, routers, lines etc.) accurately so that the results of the simulations are nearly identical with those obtained from measurements. One of the most popular tools is the Berkeley Network Simulator 
(\texttt{ns-2.26}\cite{ns}) which is used in this paper.

\begin{figure}[htb]
\begin{center}
\psfrag{0}[c][c][.45]{0}
\psfrag{1}[c][c][.45]{1}
\psfrag{i}[c][c][.45]{$i$}
\psfrag{i-1}[c][c][.45]{$i-1$}
\psfrag{N-1}[c][c][.45]{$N-1$}
\psfrag{tau}[c][c][.8]{$\tau$}
\psfrag{C}[c][c][.8]{$C$}
\psfrag{B}[c][c][.8]{$B$}
\psfrag{Xi}[c][c][.8]{$X_i$}
\psfrag{keplet}[c][c][.8]{$C_{i-1}=\sum\limits^{N-1}_{j=0, j\neq i}X_j $}
\psfrag{label}[c][c][.8]{$X_i=\frac{w_i P}{T_{RTT}}$}
\includegraphics[width=.24\textwidth]{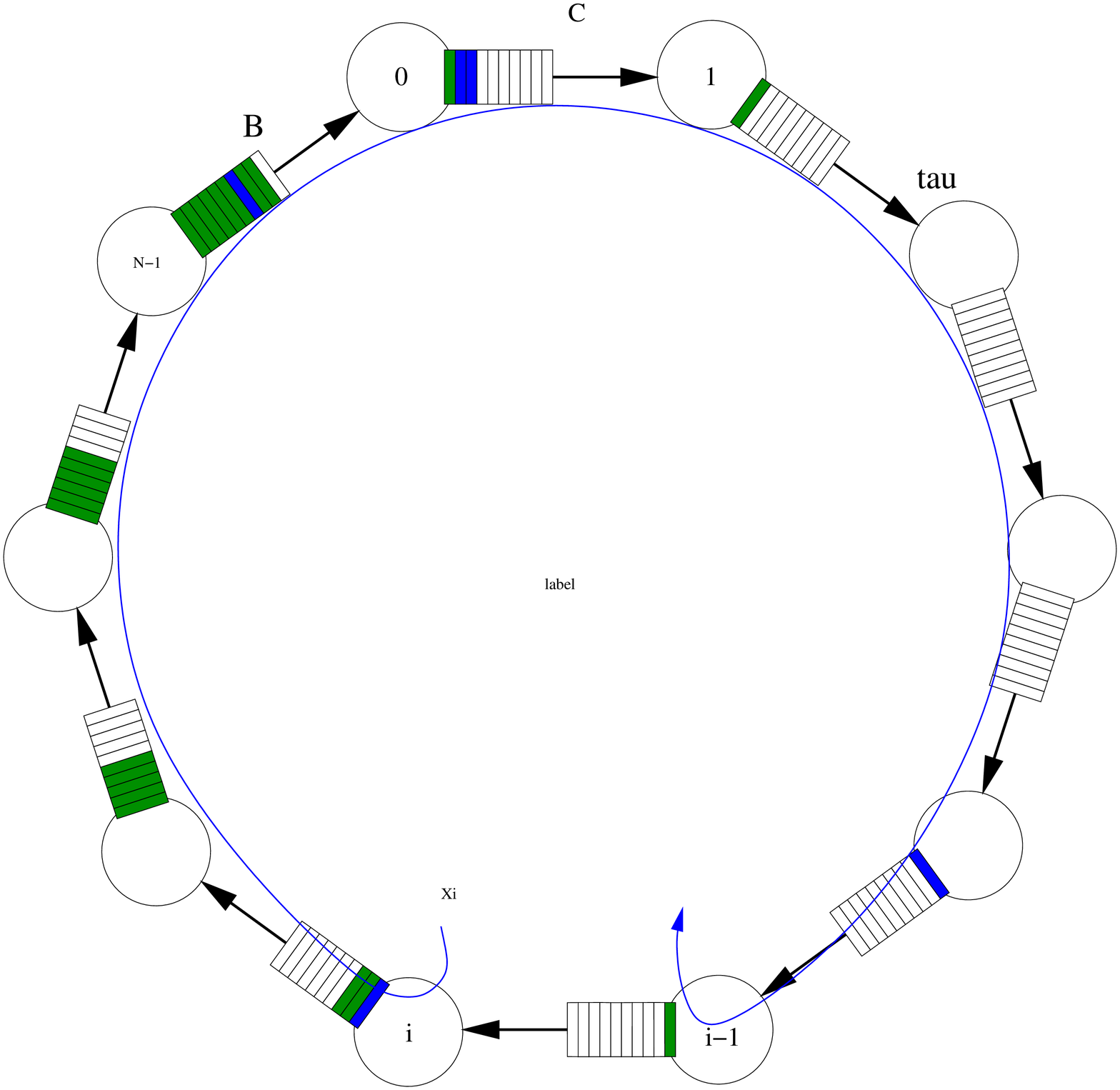}
\includegraphics[width=.22\textwidth]{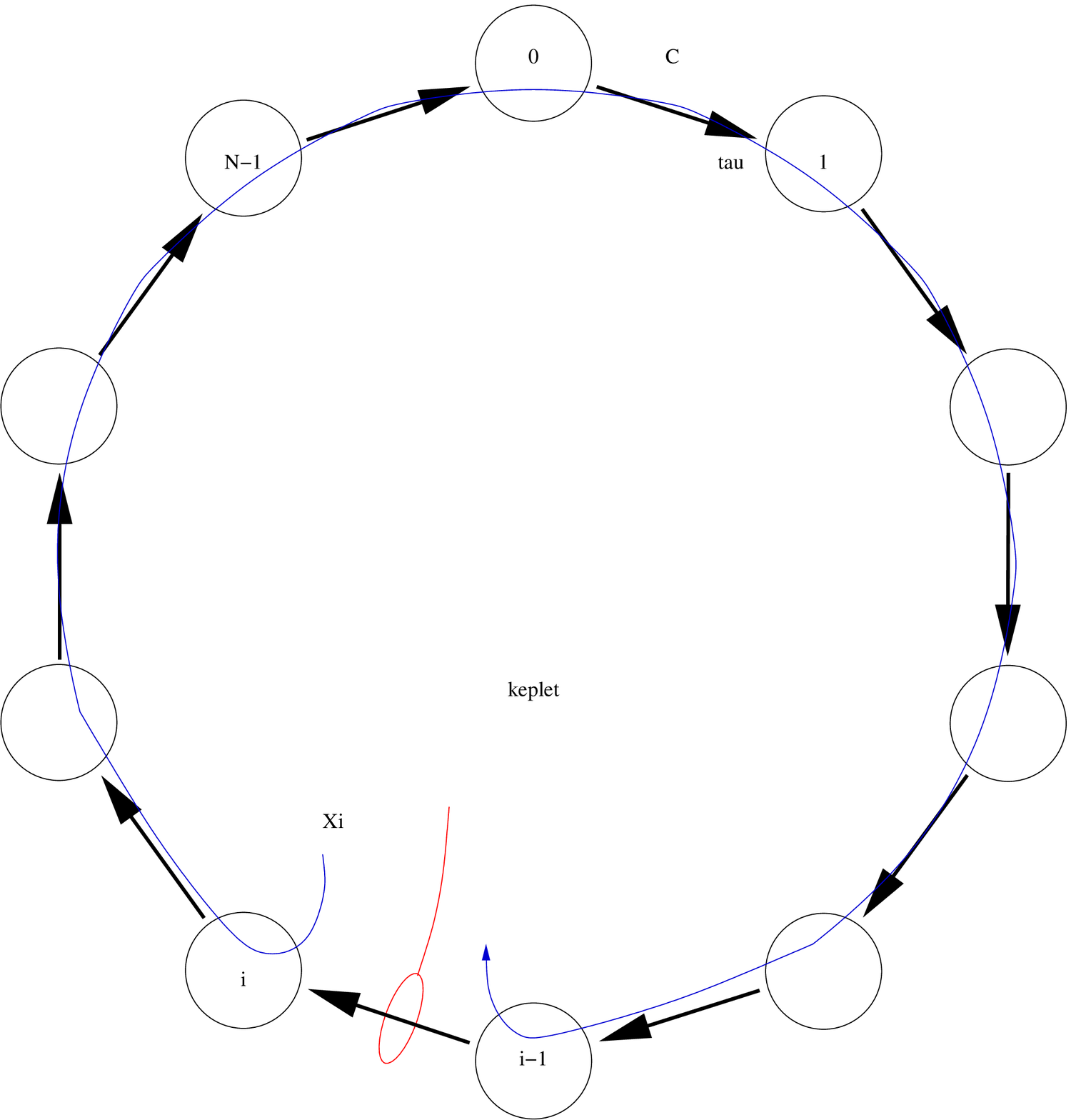}
\caption{The ring structure. {\bf Left,} Network simulator: The TCP agent at site $i$ continuously transfers data trough the ring of network routers connected with constant capacity $C$ lines with time delay $\tau$. The packets not processed yet by the routers are queued in buffers of size $B$. The stream of data can traverse in clockwise direction only to arrive at the destination node at site $i-1$. In the simulations a realistic set of parameters has been set ($C=10^7~$bits/s, $\tau=.031~$s, $B=300~$packets, $P=4416~$bits, $N=10$). {\bf Right,}Continuous model: Each TCP is modeled with its sending rate $X_i$ and its traffic flow traverses the ring topology. On the link connecting sites $i-1$ and $i$ all the sending rates of traversing TCP flows are summed up to yield the total link utilization.% $C_{i-1}=\sum\limits^{N-1}_{j=0}X_j - X_i$.
\label{topology}}
\end{center}
\end{figure}

In this letter we study the network traffic generated in a unidirectional ring of identical routers connected as shown in the left part of Fig.~\ref{topology}.
The ring geometry mimics periodic boundary conditions. This way we can study the propagation of congestion in an isolated, clean setup, which makes it possible to compare the results to those obtained in cellular automaton, car and granular traffic simulations. We show that this system drives itself into a critical congested state in a self-organized way. Both the position of the congested router and the packet sending activity at the sites propagate against the direction of the packet flow. The profile of the congestion wave can be reconstructed from the activities of the computers connected to the ring. The velocity of the congestion wave can be measured. Then a simple model of the system is introduced which is able to explain the main features of the congestion wave such as shape and speed. A formula for the velocity of the congestion wave is obtained and tested.

In our model system the ring is formed by $N$ identical routers which can forward packets in clockwise direction. Routers are connected with a line of capacity $C$ (measured in data bits per second) with a constant forwarding delay $\tau$ (measured in seconds). The incoming data flow of a router can temporarily exceed the capacity of the outgoing line. To avoid data loss in this situation the router contains a buffer of size $B$ (measured in data packets) where packets can be stored.  The computers are instructed to send data persistently to their anti-clockwise neighbors, so that the packets traverse the longest possible route in the ring. The traffic is "granular" as computers send in data packets of size $P$ (measured in bits).  The data traffic of computers is controlled by the TCP/IP protocol\cite{jacobson}. This protocol ensures that the data packet-sending rate is decreased whenever congestion occurs and that it is increased when there is an available unused capacity in the system. The control algorithm became quite sophisticated over the years. Here we can only sketch the main properties of the most popular version used currently.
  
After establishing connection between two computers over the network TCP algorithm regulates the packet-sending rate. First a single packet is sent out. Upon receiving that packet the receiver acknowledges the arrival of the packet by sending back a small size acknowledgement packet (ACK). The time elapsed between the sending out of a packet and receiving the corresponding ACK is called round trip time (RTT).  The TCP maintains an internal variable, the Congestion Window ($w$), which is used to control the number of packets sent out when the ACK is received. It starts with the initial value $w=1$ and then it is increased according to $w\mapsto w+1/w$ each time an ACK is received. Two new packets are sent out if the congestion window crosses an integer value and only a single packet otherwise. This way the integer part of the window  $[w]$ gives the number of sent but not yet acknowledged packets in the network. Assuming constant RTT during this process, the congestion window and the number of packets out in the network are increased linearly in time. This process lasts until a packet is lost somewhere in the network, indicating congestion. As a response the packet-sending rate should be decreased. So, the TCP reduces the value of the congestion window $w \mapsto \beta w$ $(\beta < 1)$ and does not send out any new packets in response to ACKs until the number of still unacknowledged packets decreases to the integer part of the new (reduced) value of the congestion window. After that the packet-sending algorithm returns to the original linear increase phase described above. 

In case the congestion window variable is large we can neglect its granularity and can treat it as a continuous variable. In the linear increase phase the window variable is increased by one in each RTT period and its dynamics can be well approximated with the differential equation 
\begin{equation}
\frac{dw}{dt}=\frac{1}{T_{RTT}(t)},
\end{equation}
where $T_{RTT}(t)$ is the actual value of the RTT measured by the TCP. In case of packet loss the window is reduced
\begin{equation}
w(t_{+})=\beta w(t_{-}),
\end{equation}
where $t_{\pm}=t\pm0$ denotes the time right after and before the packet loss. The sending rate $X$ (bit/sec) in the continuous approximation can be estimated as the amount of data sent within an RTT, $X=Pw/T_{RTT}$. Neglecting the change of RTT on the scale of RTTs the sending rate satisfies the following pair of equations:
\begin{eqnarray}
\frac{dX}{dt}&=&\frac{P}{T_{RTT}^2(t)},\label{xx}\\
X(t_{+})&=&\beta X(t_{-}) \;\;\; \mbox{at packet loss},
\end{eqnarray}
where typically $\beta=1/2$. 

\begin{figure}[htb]
\begin{center}
\includegraphics[width=\mywa]{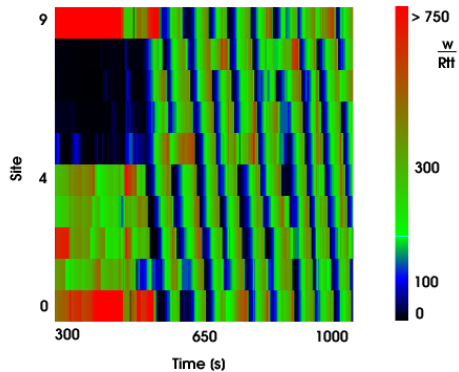}
\caption{Spatiotemporal diagram of congestion propagation. The horizontal axis is the real time (in seconds) of the system and the site index ($i$) is on the vertical axis. In this simulation the number of sites was $N=10$. Notice, that the sites $i=0$ and $i=N-1$ are neighbors in the ring topology. The sending rate $X_i$ is color-coded according to the scale indicated at the right side of the figure. Deep blue indicates very low sending rate due to high congestion. It can be seen that the most congested site propagates in anti-clockwise direction with almost constant speed, while the packet traffic itself is clockwise directed.  
\label{spatio}}
\end{center}
\end{figure}

Next, we present the results of our simulation study carried out with the network simulator. The geometry and parameters of the setup is on the left part of Fig.~\ref{topology}, which is essentially the same as we could have observed in real hardware realization. Fig.~\ref{spatio} shows the spatiotemporal diagram of the congestion wave occurring in the network simulator. One can see that after a short initial transient (up to $500~$s) the pattern remains stable and propagates in anti-clockwise direction. In this respect it resembles the congestion propagation in car traffic. The speed of the congestion wave pattern is almost constant. Its average can be determined by measuring the average speed of the center of mass of the pattern. This should be carefully defined in the present situation as the system is spatially periodic. Taken into account the periodic boundary conditions the center of mass of the pattern can be defined by $\langle{} i \rangle(t)=\arg\left(\Sigma_{j=0}^{N-1} X_j(t) e^{\img (2\pi/N) j}\right)$. The speed of the pattern is the time derivative of this quantity. 

\begin{figure}[htb]
\begin{center}
\psfrag{Site}[c][c][.8]{Site}
\psfrag{Average bandwidth occupied [bps]}[c][c][.8]{Time averaged sending rate [bits/s] }
\psfrag{Data points of the network simulator}[r][r][.7]{}
\includegraphics[width=\myw,angle=270.]{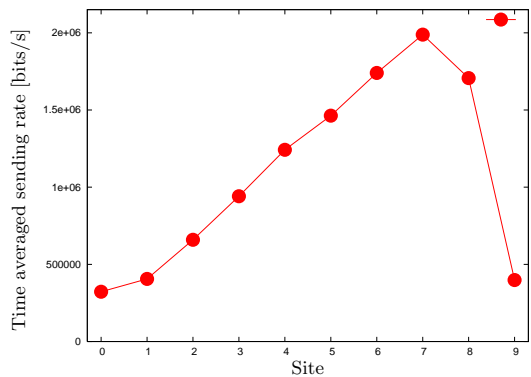}
\caption{The shape of the traveling wave profile. It has been determined by averaging the time series in co-moving spatial coordinates. 
\label{front}}
\end{center}
\end{figure}

Once the speed of the pattern is determined we can analyze the shape of the profile. Representing the sending rates $X_{i'+\lfloor\langle{}i \rangle\rfloor}(t)$ in co-moving coordinates $i'$ relative to the center of mass we recover  the shape of the traveling wave pattern.  Averaging the new series in time the profile of the front emerges as in Fig.~\ref{front}.

While the continuous equations constitute gross simplification of the original TCP dynamics, the main properties of the traveling wave can be recovered from them with some additional assumption made on the packet loss process as we show next.

The bandwidth $C_{i-1}(t)$ utilized on the link connecting nodes $i-1$ and $i$ is the sum of sending rates of TCPs whose traffic flows through that link. In our case the flows of all TCPs traverse that link except the one starting at node $i$ and ending at node $i-1$:
\begin{equation}
C_{i-1}(t)=\sum_{j=0, j\neq i}^{N-1} X_j(t) =\sum_{j=0}^{N-1} X_j(t)-X_i(t),
\label{modell} 
\end{equation}
where site $i=N$ is identified with site $i=0$ due to periodicity. The traffic of ACK packets emanating in $i-1$ and absorbed in $i$ is low due to the small size of ACK packets and its contribution to traffic can be neglected (see also the right part of Fig.~\ref{topology}). Congestion and packet loss occur in the system whenever one of the bandwidths $C_i(t)$ reaches the link capacity $C$. According to (\ref{modell}) the largest link utilization $C_i(t)$ is at site $i=i^*-1$ where $i^*$ is the site where the sending rate $X_{i^*}(t)$ is the lowest. We then have to investigate which TCP flow will lose packet on link $i^*-1$. In principle all the TCP flows traversing the congested link can lose packets, so only the TCP at site $i^*$ is immune. Our observation is that the TCP flow starting at the actual congested link (with sending rate $X_{i^*-1}$) experiences the packet loss almost surely. This is due to the fact that the processing of the small acknowledgement packets is faster than that of data packets and they tend to queue up and to arrive in batches. As a response, TCPs send out data packets also in batches. Then obviously the TCP that ejects this burst of data packets directly into an almost saturated buffer will lose packets in the process.

The mechanism described above is responsible for the emergence of the congestion wave in the system. The TCP at site $i^*-1$ suffers packet losses repeatedly until its sending rate $X_{i^*-1}$ becomes smaller than $X_{i^*}$. From then on $X_{i^*-1}$ will be the lowest in the system, link utilization $C_{i^*-2}$ will be the highest and TCP at site $i^*-2$ suffers the packet losses. This way congestion propagates site by site anticlockwise in the system.  After several rounds of congestion propagation the propagating front of Fig.\ref{front} emerges. 

\begin{figure}[htb]
\begin{center}
\psfrag{1}[c][c][.9]{0}
\psfrag{2}[c][c][.9]{1}
\psfrag{N-1}[c][c][.9]{$N-1$}
\psfrag{a}[r][l][.9]{$a$}
\psfrag{a+b}[r][l][.9]{$a+b$}
\psfrag{X}[c][c][.9]{$X$}
\psfrag{i}[c][c][.9]{$i$}
\includegraphics[width=\myw]{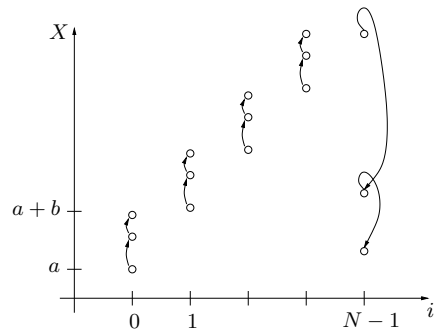}
\caption{Sending rate evolution in time. $N=5$, $l=2$, $X_4$ decreases until
it reaches the value of $X_0$.
\label{explain}}
\end{center}
\end{figure}

The shape of the front is linear with a sharp drop connecting its ends. We can determine the parameters of the linear front in our model.   Let the nodes forming the linear part of the front range from $0$ to $N-1$ and let the minimum sending rate first be at the $0$th node 
\begin{equation}
X_i(0)=a+bi,
\label{linear}
\end{equation}
as it is shown in Fig.~\ref{explain} that illustrates the evolution of a system containing $N=5$ nodes.
The packets start to be dropped at site $N-1$. We start our description at the first packet drop, which occurs when $C_{N-1}(0)=\sum_{i=0}^{N-1}X_i(0)-X_0(0)=C$ holds. This initial condition gives the first condition for the initial shape of the front
\begin{equation}
C=(N-1)\left(a+\frac{N}{2}b\right).
\label{limit}
\end{equation}
Immediately after the packet drop the sending rate at site $N-1$ decreases to $X_{N-1}(+0)=\beta X_{N-1}(0)$, while the rest of the sending rates stay unchanged. Then the sending rates increase in a uniform manner with an amount ${X'}_i=X_i+(1-\beta)X_{N-1}/(N-1)$ until the next packet loss occurs. In particular, from before the first packet loss until the second packet loss the sending rate at site $N-1$ changes to ${X'}_{N-1}=qX_{N-1}(0)$, where   $q=\beta+\frac{1-\beta}{N-1}$. This process is then repeated $l$ times until the sending rate $q^l X_{N-1}$ becomes lower than the actual value of $X_0$.  This way the wave moves one site to the left, while its linear shape is preserved as the sending rates of all nodes, except $N-1$, increase with $b$. From Eq.~\ref{xx} one can calculate the time needed for this $T_p=bT_{RTT}^2/P$. Accordingly the speed of 
the congestion wave is $v:=1/T_p$ (measured in site/sec.). 

\begin{figure}[ht]
\begin{center}
\includegraphics[width=\myw, angle=270]{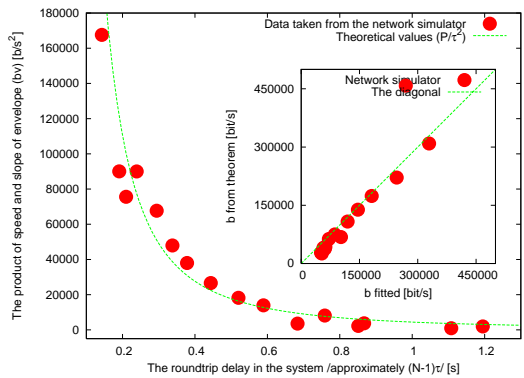}
\caption{
Testing the relation $vbT_{RTT}^2/P= 1$  for various system sizes $N=5-20$ in the network simulator. After the end of transients the values of $b$, $T_{RTT}$ and $v$ have been obtained
by averaging over samples of $5000$ sec.  {\bf Inset:}
Testing Eq.~\ref{resultslope} in the network simulator. Average values of $b$ are plotted against
the results of (\ref{resultslope}) with $q^l$ obtained from the simulation.
\label{tests}}
\end{center}
\end{figure}

The formula derived for the speed of the congestion waves can be tested against real data produced by the network simulator. In the simulation one can measure the average round trip time observed by the TCPs, the mean slope of the linear part of the front $b$ and the speed of the front.  On the main part of Fig.~\ref{tests} one can see that the measured values satisfy the relation $vbT_{RTT}^2/P= 1$ very well. 

With the help of  (\ref{limit}) and (\ref{linear}) the slope of  the front in the model can also be directly expressed as
\begin{equation}
b=\frac{2C(1-q^l)}{(N-1)(N-2)(q^l+1)},\label{resultslope}
\end{equation}
where $l$ should be determined independently. Our mathematical model allows several positive integer values of $l$ with an upper bound due to the positiveness of $b$.  We found that only the largest possible $l$ value is stable against small perturbations of the wave front. Systems started at lower $l$ values always shift towards a greater value of  $l$. In the network simulator we always observed the realization of the most stable (the highest possible $l$) solutions of the model. In the inset of Fig.~\ref{tests} we compare the measured values of $b$  with (\ref{resultslope}). We again find good agreement.

As a summary we showed that congestion waves are formed naturally in the data traffic of today's computer networks. The mechanism behind the wave formation is that packet losses occur most likely in computers nearest to the site of the actual congestion and other computers sharing the congested link increase their sending rates, moving the site of the congestion one site downstream. This basic mechanism is quite general and can create congestion moving against the direction of the data traffic in more complicated geometries. A formula for the speed of the congestion wave has been derived in a simple ring topology and network simulations have confirmed it. Such formulas can be developed for more complicated geometries, which is our next research goal.      

The authors thank the National Science Fund, Hungary (OTKA T37903) for the support.

\end{document}